\documentclass[twocolumn,prb,showpacs,superscriptaddress]{revtex4}
\usepackage{graphicx,psfrag,amssymb,amsmath,color}
\begin{document}
\title{Pseudogap enhancement due to magnetic impurities in d-density waves}
\author{Bal\'azs D\'ora}
\email{dora@kapica.phy.bme.hu}
\affiliation{Department of Physics, Budapest University of Technology and
Economics, H-1521 Budapest, Hungary}
\author{Kazumi Maki}
\affiliation{Department of Physics and Astronomy, University of Southern
California, Los Angeles CA 90089-0484, USA}
\author{Attila Virosztek}
\affiliation{Department of Physics, Budapest University of Technology and
Economics, H-1521 Budapest, Hungary}
\affiliation{Research Institute for Solid State Physics and Optics, P.O.Box
49, H-1525 Budapest, Hungary}
\author{Andr\'as V\'anyolos}
\affiliation{Department of Physics, Budapest University of Technology and 
Economics, H-1521 Budapest, Hungary}

\date{\today}

\begin{abstract}
We study the effect of quantum magnetic impurities on d-wave spin density waves (d-SDW). The impurity spins
are aligned coherently according to the spin space anisotropy of the condensate.  
Both the order parameter and transition temperature increases due to the coherent interplay between 
magnetic scatterers and d-SDW. This can explain the recent experimental data on the pseudogap enhancement of 
Ni substituted NdBa$_2$$\{$Cu$_{1-y}$Ni$_y\}$O$_{6.8}$ from Pimenov et al. (Phys. Rev. Lett. \textbf{94}, 227003 
(2005)).

\end{abstract}


\maketitle


The nature of the pseudogap phase of high $T_c$ superconductors is highly controversial. Among many others, Chakravarty 
et al.\cite{nayak} proposed that it results from d-density wave order. This phase is a charge density wave 
(CDW)\cite{gruner} with a gap of d-wave symmetry. This causes the lack of periodic charge modulation, and evokes the 
notion "hidden-order". However, there are many other unconventional density wave phases\cite{Ozaki}, with similar 
properties to d-density waves\cite{cappelluti, benfatto,valenzuela}. For example, d-wave spin density waves (d-SDW) are 
almost 
indentical to d-CDW except the 
hidden spin space anisotropy, and can account for several experimental results in the pseudogap phase of high $T_c$ 
superconductors equally well as d-CDW does.

Last year, Pimenov et al.\cite{pimenov} reported that the pseudogap energy in underdoped  
NdBa$_2$CuO$_{6.8}$ is enhanced by substituting Cu in the CuO$_2$ plane with Ni by measuring the $c$-axis optical conductivity. We have established recently that the 
pseudogap phase should be d-wave density wave through the analysis of both the giant Nernst effect and the angle 
dependent magnetoresistance\cite{capnernst,dorauj}. However, we cannot decide whether it is d-wave charge or spin 
density wave. The earlier 
Pauli limiting behaviour of d-density waves suggested d-CDW\cite{elbaum1,elbaum2,elbaum3}. However, it is easily seen 
that d-SDW 
also exhibits Pauli 
limiting when the spin anisotropy axis in d-SDW lies in the $a-b$ plane. Further a recent analysis of impurity 
scattering\cite{vanyimp} under general conditions suggests that such an enhancement of d-density wave order is 
impossible, if the 
condensate is d-CDW and/or the impurity is nonmagnetic\cite{ghosal,yang}. This leaves us the unique possibility that 
d-density wave needs 
to be d-SDW and the impurity scattering is of Kondo type.

Therefore we consider in the followings d-SDW in the presence of magnetic impurities with the Kondo coupling. The 
study of magnetic imputities has a long history\cite{magnabrikosov,shiba}. In order 
to couple the impurity spins to the underlying d-SDW, the Kondo coupling has to be non-local and has to have a d-wave 
component as in Ref. \onlinecite{roma}.
The interaction between quantum magnetic impurities and 
d-SDW is similar in essence to the RKKY interaction, but the electrons mediating the coupling between the spins are not 
free but participate in collective phenomenon, and belong to the density wave condensate.
This leads to the enhancement of the pseudogap energy, similarly to 
ferromagnetic superconductors\cite{jaccarino}. 
There, the Pauli term due to an external magnetic field
can compensate the exchange term due to ferromagnetism, and superconductivity will be revived. 
It is worth mentioning that d-density waves have unusual magnetic properties 
with the inclusion of spin-orbit coupling as well\cite{wu}.

In general, nonmagnetic impurities such as Zn, lead to the destruction of the underlying phase\cite{pan,hudson}, 
regardless to 
its symmetry, and are thought to belong to the unitary scattering limit. Ni impurities have also similar impact on 
superconductors due to the lack of spin space anisotropy, and 
act as simple potential scatterers in the Born limit. In the situation considered below, their magnetism can couple 
directly to the order parameter of d-SDW, completely modifying the simple scattering picture. No such mechanism is 
expected to exist in superconductors, hence magnetic Ni impurities suppress superconductivity similarly to Zn.

\begin{figure}[h!]
\centering
{\includegraphics[width=8.5cm,height=2cm]{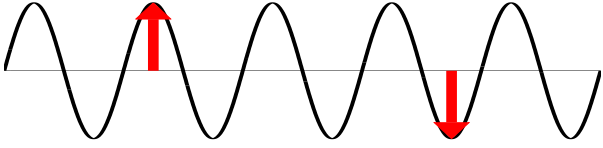}}
\caption{(Color online) The impurity spin configuration is sketched in the ordered phase. 
The solid line denotes the alternating field dictated by the impurity-density wave interaction, 
the colored objects represent the impurity spins.
\label{config}}
\end{figure}

As a model, we consider an effective low-energy Hamiltonian describing d-wave spin density waves (d-SDW)\cite{IO,nagycikk}, interacting with
randomly distributed quantum magnetic impurities:
\begin{gather}
H=\sum_{\bf k,\sigma}\left[\varepsilon({\bf k})a^+_{\bf k,\sigma}a_{\bf k,\sigma}+\textmd{i}\Delta\sigma f({\bf 
k})a^+_{\bf 
k,\sigma}a_{\bf k-Q,\sigma}\right]+\nonumber\\
+\sum_{{\bf r,R}_j}J({\bf r-R}_j){\bf S}_j{\bf s(r)}.
\end{gather}
Here $\varepsilon({\bf k})$ is the kinetic energy spectrum, $\Delta$ is the d-SDW order parameter, $f({\bf 
k})=(\cos(k_xa)-\cos(k_ya))/2$ is the gap function with d-wave symmetry.
For simplicity, and to make connection with earlier analysis\cite{nayak}, we study a commensurate situation with best 
nesting vector 
${\bf Q}=(\pi/a,\pi/a)$. A magnetic impurity (${\bf S}_j$) at ${\bf R}_j$ interacts with the electron spin density ($\bf 
s(r)$) 
through the 
non-local Heisenberg exchange coupling. 
Its matrix element\cite{roma} describing scattering with $\bf Q$ is expanded in terms of Fermi surface harmonics, 
and we retain 
the term possessing d-wave symmetry, namely $J({\bf k,Q})\propto f({\bf k})$, the others are unable to couple to the wavevector dependence of the gap. 
To investigate the effect of magnetic interaction on d-SDW, we study its change on the grand canonical potential 
perturbatively. To lowest order, one finds
\begin{equation}
\Delta\Omega=-\frac{2J\Delta}{P}\sum_{{\bf R}_j}S_j^z\cos({\bf QR}_j),
\end{equation}
which suggests that each impurity feels a local aligning "magnetic" field manifested through $\Delta/P\cos({\bf 
QR}_j)$,  $S^z$ is the component of the impurity spin parallel to the spin of underlying d-SDW which lies in
the $a-b$ plane.
Depending on the sign of $J\Delta$, the impurity will order locally in ferro- or antiferromagnetic fashion with respect 
to the local field\cite{roma}, but the 
overall arrangement of impurities will follow an antiferromagnetic pattern due to the $\cos({\bf QR}_j)$ factor, as is 
visualized schematically in Fig. \ref{config}.  

\begin{figure}[h!]
\centering
\psfrag{x}[t][b][1.2][0]{$n_iJS/\Delta_0$}
\psfrag{y}[b][t][1.2][0]{$\Delta/\Delta_0$, $m/S$}
{\includegraphics[width=7cm,height=7cm]{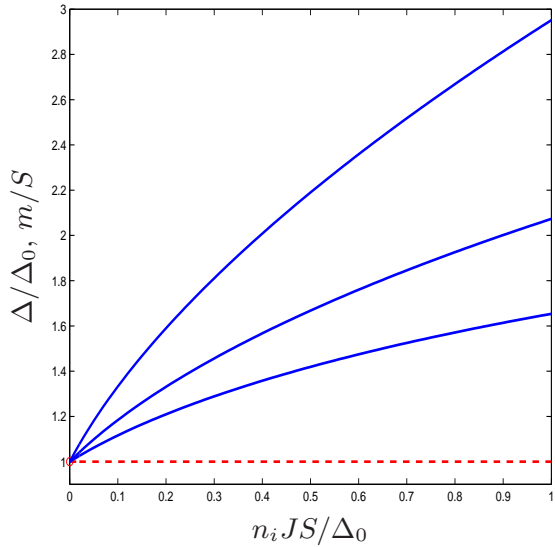}}
\caption{(Color online) The concentration and magnetic coupling dependence of the order parameters is shown 
($\Delta$ with blue solid, $m$ with
red dashed line) at $T=0$ in the weak-coupling limit for $P\rho_0=0.7$, 0.5 and 0.3 from bottom to top. Note the strong increase in $\Delta$ even for low impurity 
concentrations.
\label{deltanj0}}
\end{figure}

To make this picture more quantitative, we perform a two order parameter mean field theory, one for the d-SDW and 
another for the impurities by assuming, that the impurity spins are aligned as $S_j^z=m\cos({\bf QR}_j)$.
The mean field decoupled magnetic interaction reads as
\begin{gather}
H_{s-d}=\frac JN\sum_{{\bf k,q,\sigma,R}_j}\textmd{i}me^{\textmd{i}{\bf qR}_j}f({\bf k})\sigma a^+_{\bf 
k+q,\sigma}a_{\bf 
k-Q,\sigma}-\nonumber\\
-\frac{J}{P}2\Delta\sum_{{\bf R}_j}S_j^z+N_i\frac{Jm2\Delta}{P}.\label{sdham}
\end{gather}
Here $N$ is the number of unit cells, $P$ is the d-SDW interaction\cite{nagycikk} and $N_i$
is the number of impurity atoms.
To average over impurities, we use the standard approach to consider only non-crossing ladder type 
diagrams\cite{klasszikus}.
Usually, in the first order Born approximation, nonmagnetic impurities simply shift the chemical 
potential\cite{vanyimp}.  In our
case, however, magnetic impurities couple to the order parameter, $\Delta$, and in the first order Born approximation, 
the new energy spectrum is given by
\begin{gather}
E_{\pm}({\bf k})=\frac{\varepsilon({\bf k})+\varepsilon({\bf k-Q})}{2}\pm\nonumber\\
 \pm \sqrt{\left(\frac{\varepsilon({\bf k})-\varepsilon({\bf k-Q})}{2}\right)^{2}+f^2({\bf k})
(\Delta+n_iJm)^2}.\label{energ}
\end{gather}
The use of the Born approximation is justified from the fact that Ni impurities in unconventional superconductors 
can 
convincingly be described with it, while Zn substituted samples call for the unitary limit\cite{pan,hudson}.
The stability conditions of the order parameters can be obtained from the grand canonical potential, which takes the 
from
\begin{gather}
\Omega=N\frac{\Delta^2}{P}-T\sum_{\begin{array}{c} {\bf k}\in \textmd{RBZ} \\ 
\sigma,\alpha=\pm\end{array}}\ln\left(1+e^{\beta(E_\alpha({\bf 
k})-\mu)}\right)+\nonumber\\
+\frac{Jm}{P}N_i2\Delta-TN_i\ln\left(\dfrac{\sinh\left(\beta J\Delta(2S+1)/P\right)}{\sinh\left(\beta 
J\Delta/P\right)}\right).
\end{gather}
Here $S$ is the impurity spin quantum 
number, RBZ stands for the reduced (antiferromagnetic) Brillouin zone.
The first two terms are characteristic to density waves\cite{gruner}, the third one stands for the electron-impurity interaction, 
while the latter describes to certain extent the quantum nature of the magnetic impurity, as opposed to classical 
spins\cite{shiba}. After minimization with 
respect to $\Delta$ and $m$, the coupled gap equations are obtained as
\begin{gather}
\Delta=\frac{2P}{N}\sum_{{\bf k}\in \textmd{RBZ}}\left(f(E_{-}({\bf k})\right)-f\left(E_{+}({\bf 
k})\right)\frac{(\Delta+n_iJm)f^2({\bf k})}{E_+({\bf k})-E_{-}({\bf k})},\label{gapdelta} \\
2m=(2S+1)\coth\left(\frac{\beta J\Delta(2S+1)}{P}\right)-\coth\left(\frac{\beta J\Delta}{P}\right),\label{gapm}
\end{gather}
$n_i=N_i/N$, $f(x)$ is the Fermi distribution function.
The first one is the usual BCS type gap equation, while the second one with Brillouin function on the right hand side 
accounts for the quantum nature of the magnetic impurity.
Regardless to the detailed form of the kinetic energy spectrum, Eq. \eqref{gapm} can be solved at $T=0$ to give 
$m=S$, the impurity spin is aligned  completely. The above gap equations can further be simplified in the continuum 
limit\cite{valenzuela} and 
assuming perfect nesting ($\varepsilon({\bf k})+\varepsilon({\bf k-Q})=0$), but we would like to emphasize that our 
result are robust with respect to variation of the kinetic energy spectrum, commensurability and imperfect nesting.
With this, the zero temperature d-SDW order parameter is determined from
\begin{equation}
\frac{2n_iJS}{P\rho_0}=(\Delta+n_iJS)\ln\left(\frac{\Delta+n_iJS}{\Delta_0}\right)
\end{equation}
with $\rho_0$ the normal state density of states, $\Delta_0$ is the order parameter of the pure system at $T=0$.
In spite of its relative 
simplicity, this formula constitutes the main result of this paper and is shown in Fig. \ref{deltanj0} for various 
interactions. For any finite 
$n_iJ$, 
$\Delta$ is enhanced with 
respect to its pure value $\Delta_0$. The steady increase stems from Eq. \eqref{sdham}, since impurities act like a 
source term. These findings are in accordance with Ref. \onlinecite{pimenov}, where a steady 
increase of the 
pseudogap has been reported with respect to Ni impurity concentration, by focusing on the suppression of the electric conductivity. Our result indicates that this can arise from the 
coherent interplay of magnetic impurities and d-SDW. 
Moreover, the experimentally measured pseudogap energy scale is not simply $\Delta$, but $\Delta+n_iJS$ as seen from 
Eq. \eqref{energ}, causing further enhancement of the pseudogap. 
As opposed to this, Zn impurities act like potential scatterers, and provide us with pair-breaking effect, leading to 
the destruction of the superconducting or density wave condensate.

\begin{figure}[h!]
\centering
\psfrag{x}[t][b][1.2][0]{$T/T_{c0}$}
\psfrag{y}[b][t][1.2][0]{$\Delta/\Delta_0$, $m$}
{\includegraphics[width=7cm,height=7cm]{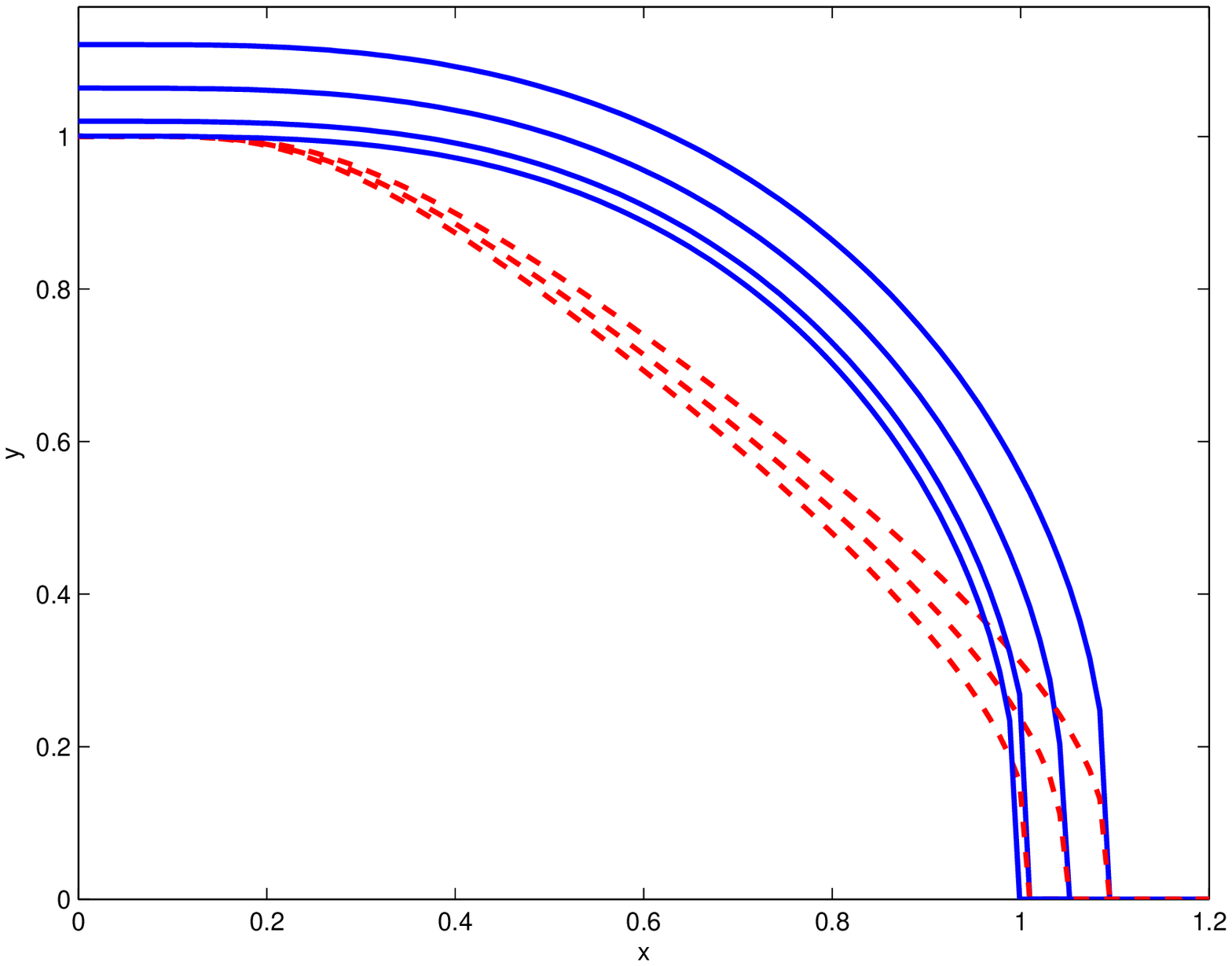}}
\vspace*{1cm}

{\includegraphics[width=7cm,height=7cm]{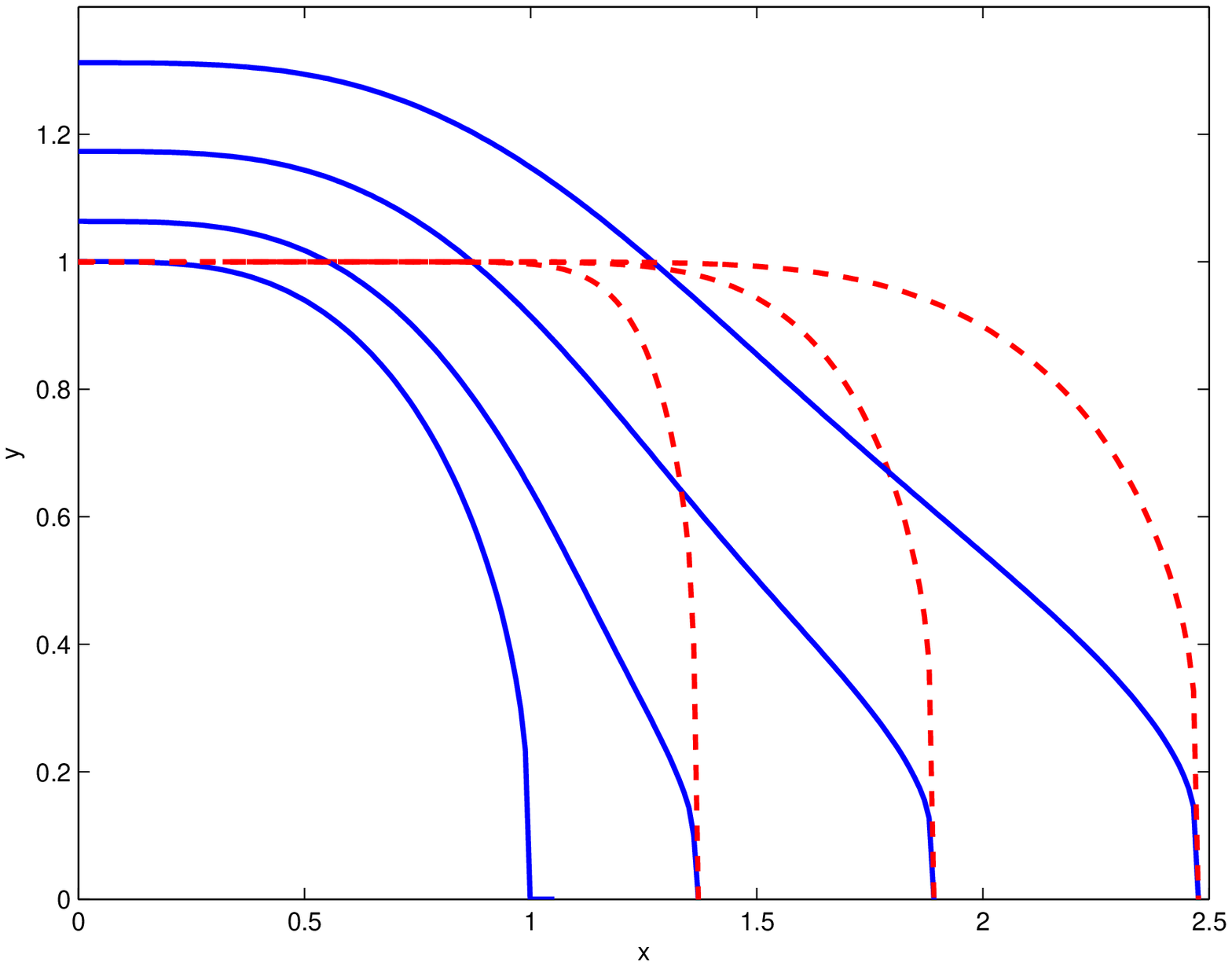}}
\caption{(Color online) The temperature dependence of the order parameters is shown ($\Delta$ with blue solid, $m$ with
red dashed line) in the weak-coupling limit for
$P\rho_0=0.5$ and $S=1$. The top panel shows weak impurity-electron coupling $\Delta_0/J=4.5$  with $n_i=0$, 0.03,
0.1 and 0.2 with increasing $T_c$. In the bottom panel, strong impurity-electron coupling is used with 
$\Delta_0/J=0.45$,
$n_i=0$, 0.01, 0.03 and 0.06 with increasing transition temperature. For stronger coupling, small concentration causes
dramatic increase in both $\Delta$ and $T_c$.
\label{deltajbig}}
\end{figure}

The change in the transition temperature ($T_c$) is different from the Abrikosov-Gor'kov formula deduced for 
non-magnetic 
impurities, and reads as
\begin{equation}
\ln\left(\frac{T_c}{T_{c0}}\right)=\frac{1}{P\rho_0}\frac{2n_iJ^2S(S+1)}{3PT_c+2n_iJ^2S(S+1)}.
\end{equation}
The right hand side of this equation is always positive, hence similarly to $\Delta$, $T_c$ also enhances in the 
presence of magnetic impurities. For arbitrary temperatures, the 
gap-equations have been solved numerically, and are shown in Fig. \ref{deltajbig} for $S=1$, characteristic to Ni. When 
the Heisenberg coupling ($J$) is 
large, the spins retain their maximum value ($S$) up until the close vicinity of $T_c$, and a small amount of 
impurities increase significantly $\Delta$. On the other hand, for weaker $J$, the d-SDW order parameter follows the 
usual BCS like temperature dependence, and $m$ reaches its maximum value only at low temperatures.

We mention here, that 
the finite value of $m$ results in small amplitude antiferromagnetic modulation of the localized spins. Recently 
such weak low temperature staggered magnetization lying in the $a-b$ plane has been observed in 
YBa$_2$Cu$_3$O$_{6.5}$\cite{sidis} with an amplitude $m_0=0.05\mu_B$ well above the superconducting transition 
temperature (55~K) at 310~K. We believe that this antiferromagnetism is related to the interaction between the 
pseudogap phase and localized moments. Indeed, the measured magnetic intensity closely resembles to that in Fig. 
\ref{deltajbig}. Albeit orbital antiferromagnetism\cite{nayak} does not exist in our d-SDW model as opposed to its d-CDW counterpart, the sensitivity of its order parameter to local magnetic fields provides us with weak antiferromagnetism, similarly to URu$_2$Si$_2$\cite{roma}.

Similar phenomenon is expected to occur in itinerant antiferromagnets (namely conventional spin density waves with constant 
order parameter\cite{gruner}) interacting with magnetic impurities. The impurities will align following the spin density oscillations of 
electrons, leading to an enhancement of the ordering amplitude. We predict the realization of the aforementioned coherent 
interplay between magnetic impurities and condensate in the SDW phase of Bechgaard salts (TMTSF)$_2$X with X=PF$_6$, 
ClO$_4$, AsF$_6$ etc. polluted with magnetic 
scatterers.

In conclusion, we have demonstrated that magnetic impurities are coherently aligned in a d-wave spin density wave, 
causing the increase of the order parameter of the condensate. This can be identified with the pseudogap energy scale, 
whose enhancement is in agreement with a recent experiment in Ni substituted 
NdBa$_2$$\{$Cu$_{1-y}$Ni$_y\}$O$_{6.8}$\cite{pimenov}. As was pointed out, magnetic correlations play an important role 
in the explanation. We predict, that beyond the pseudogap enhancement in 
NdBa$_2$$\{$Cu$_{1-y}$Ni$_y\}$O$_{6.8}$, small amplitude antiferromagnetic ordering should be observable due to the Ni 
impurities, similarly to YBa$_2$Cu$_3$O$_{6.5}$\cite{sidis}.

\begin{acknowledgments}
We thank Christian Bernhard and Bernhard Keimer for very useful discussions on pseudogap enhancement, which motivated 
the present study.
B. D\'ora was supported by the Magyary Zolt\'an postdoctoral
program of Magyary Zolt\'an Foundation for Higher Education (MZFK).
This work was supported by the Hungarian
Scientific Research Fund under grant numbers OTKA TS049881, T046269 and NDF45172.
\end{acknowledgments}

\bibliographystyle{apsrev}
\bibliography{nini}
\end{document}